\documentclass{emulateapj}
\usepackage{apjfonts}
\usepackage{epsfig}
\usepackage{natbib}

\newcommand{\msun}{\ensuremath{M_{\odot}}}

\slugcomment{To Appear in {\it The Astrophysical Journal}}
\shorttitle{X-rays and Dust Emission from Kes 75}
\shortauthors{Morton et al.}

\begin{document}

\title{Observations of X-rays and Thermal Dust Emission from the Supernova 
Remnant Kes~75}

\author{Timothy D. Morton\altaffilmark{1}, Patrick Slane\altaffilmark{2},
Kazimierz J. Borkowski\altaffilmark{3}, Stephen P.
Reynolds\altaffilmark{3}, David J. Helfand\altaffilmark{4}, 
B.~M.~Gaensler\altaffilmark{5,2}, and John~P.~Hughes\altaffilmark{6}}
\altaffiltext{1}{Harvard College, Cambridge, MA 02138.}
\altaffiltext{2}{Harvard-Smithsonian Center for Astrophysics, 60 Garden
Street, Cambridge, MA 02138.}
\altaffiltext{3}{Department of Physics, North Carolina State University, Box
8202, Raleigh, NC 27695-8202.}
\altaffiltext{4}{Columbia Astrophysics Laboratory, Columbia University, 550
West 120th Street, New York, NY 10027.}
\altaffiltext{5}{School of Physics A29, The University of 
Sydney, NSW 2006, Australia.}
\altaffiltext{6}{Department of Physics and Astronomy, Rutgers University, 136
Frelinghuysen Road, Piscataway, NJ 08854.}

\begin{abstract}

We present {\sl Spitzer Space Telescope} and {\sl Chandra X-ray
Observatory} observations of the composite Galactic supernova remnant
Kes 75 (G29.7$-$0.3). We use the detected flux at 24 $\micron$ and hot
gas parameters from fitting spectra from new, deep X-ray observations
to constrain models of dust emission, obtaining a dust-to-gas mass
ratio $M_{dust}$/$M_{gas}\sim$10$^{-3}$. We find that a two-component
thermal model, nominally representing shocked swept-up interstellar or
circumstellar material and reverse-shocked ejecta, adequately fits the
X-ray spectrum, albeit with somewhat high implied densities for both
components. We surmise that this model implies a Wolf-Rayet progenitor
for the remnant. We also present infrared flux upper limits for the
central pulsar wind nebula.

\end{abstract}


\section{Introduction}

The Galactic supernova remnant (SNR) Kes~75, together with its associated
pulsar (PSR J1846-0258) and pulsar wind nebula (PWN), is a prototypical
example of a composite supernova remnant. Located at a distance of $d
\sim 19 d_{19}$~kpc (based on neutral hydrogen absorption measurements
-- Becker \& Helfand 1984) the pulsar luminosity is extremely high --
second only to the Crab for Galactic pulsars. PSR J1846-0258 also has an
exceptionally strong dipole magnetic field ($\sim 5\times 10^{13}$~G),
as inferred from the pulsar spin properties (Gotthelf et al.~2000),
and the X-ray luminosity from the associated PWN is a remarkably high
fraction of the pulsar's total spindown energy loss (20\%, compared to
a typical value of $\sim0.1$\%)\footnote{
$L_{x,psr}(0.5-10 {\rm\ keV}) = 4.4 \times 10^{35}{\rm\ ergs\ s}^{-1}$; 
$L_{x,neb}(0.5-10 {\rm\ keV}) = 1.7 \times
10^{36} d_{19}^2 {\rm\ ergs\ s}^{-1}$ (H03); $\dot E = 8.3 \times
10^{36} {\rm\ ergs\ s}^{-1}$ (Livingstone et al. 2006)}. The
characteristic age of the pulsar is extremely young -- only 723 years
(Gotthelf et al.~2000) -- and recent timing measurements of the braking
index, n, have yielded an estimated upper limit to its true age of only 884
years (assuming n is constant; Livingstone et al.~2006), placing it among 
the youngest known rotation-powered pulsars in the Galaxy. The
remnant is also quite large ($\sim 10$~pc in radius), implying an
average expansion velocity of $\sim$10$^4$ km/s -- typical (or even
in excess) of values for undecelerated SNR expansion.

Kes~75 has been studied in considerable detail in both the radio and X-ray
bands, but high extinction due to its large distance and location near
the Galactic plane prevent its detection at optical wavelengths.  In the
radio, a synchrotron shell is observed, with a flux density of 10 Jy at
1 GHz, and spectral index $\alpha = 0.7$ (energy flux $S_{\nu} \propto
\nu^{-\alpha}$), as well as flatter ($\alpha \sim 0.25$) emission from the
PWN (Becker \& Kundu 1976).  Observations with the {\em ASCA} observatory
reveal distinct spectral components from the PWN and the remnant shell
(Blanton \& Helfand 1996), although the SNR plasma parameters are poorly
determined due, in part, to mixing of these two spectral components as a
result of the poor angular resolution.  Similar morphology to that seen
in the radio band has been detected in the X-rays (Helfand et al. 2003 -
hereafter H03), with spatial coincidence of the thermal X-ray emission
and radio shell (see Figure 1), and an axisymmetric PWN structure with a
photon index of $\Gamma = 1.92$ (photon number flux $F_{\gamma} \propto
\nu^{-\Gamma}$).

While radio and X-ray observations provide information regarding the age,
energetics, and ambient conditions for SNRs, infrared observations of
remnants are also instructive because they can reveal the presence of
heated dust, both shock-heated swept-up dust from the circumstellar
or interstellar medium (CSM/ISM), and potentially, dust created by
the supernova itself. The Galactic Legacy Infrared Mid-Plane Survey
Extraordinaire (GLIMPSE; Benjamin et al. 2003), which uses all four bands
of the Infrared Array Camera (IRAC) on the {\sl Spitzer Space Telescope}
($SST$; 3.6, 4.5, 5.8, and 8 $\micron$), encompassed $\sim 100$ known
radio/X-ray SNRs, about 10-15\% of which were detected with confidence
(Lee et al. 2005, Reach et al. 2006). Kes~75 was not detected in the
GLIMPSE data, nor in earlier infrared SNR surveys (e.g. Arendt 1989,
Saken et al. 1992).

As part of a general {\sl SST} survey of LMC SNRs, Borkowski et al.~(2006:
B06) and Williams et al.~(2006a; W06) reported detection of four Type
Ia remnants and four core-collapse remnants, respectively, in the 24
$\mu$m and 70 $\mu$m MIPS (Multiband Imaging Photometer for Spitzer)
bands, with stringent upper limits on emission from the IRAC bands.
Using these observations in connection with dust emission models, they
find evidence of substantial dust destruction in the blast waves of both
types of remnant: 30\% -- 40\% of the total dust mass, and as much as
90\% of the mass in grains smaller than about 0.04 $\mu$m. They find
no evidence for emission associated with ejecta. Both studies report
pre-shocked dust/gas mass ratios lower by a factor of several than are
generally assumed for the LMC.

In order to explore evolutionary scenarios and investigate the dust
content in Kes~75, we present X-ray spectral fits for the remnant from
new $Chandra$ observations, as well as the first infrared detection of
this remnant from new $SST$ observations. These infrared observations have
high enough angular resolution that, in addition to determining properties
of the remnant shell, we are also able to place constraints on infrared
emission from the central PWN in order to investigate the need for the
spectral break that most PWNe exhibit between the radio and X-ray bands.

\begin{figure}[t]
\includegraphics*[width=3.3in]{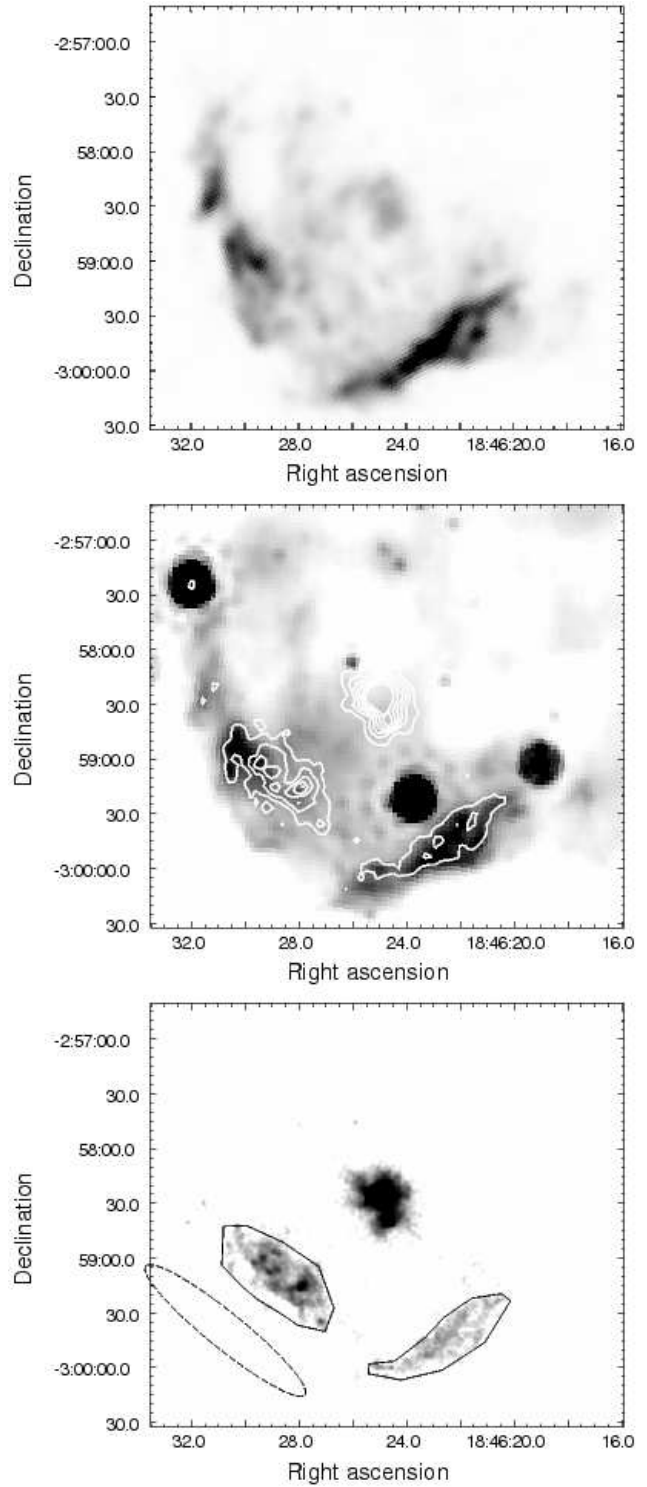}
\caption{
Top: VLA image of Kes~75 at 1.4~GHz from the MAGPIS project (Helfand et al.
2006). The greyscale is linear, covering a range of 0 to 0.1 Jy~beam$^{-1}$.
Middle: 24 $\micron$ MIPS image of Kes~75. X-ray contours are overlaid
to highlight the similar morphology. The PWN is not detected in the
infrared, nor is any part of the shell of Kes~75 detected in any of the
IRAC bands. The greyscale is logarithmic, from 100 to 200 MJy~sr$^{-1}$.
The bright, circular sources in the image, as well as numerous fainter
sources, are consistent with point sources along the line of sight to
Kes~75.
Bottom: $Chandra$ X-ray image of Kes~75. Intensity is displayed
logarithmically from 10 to 140 cnts~arcsec$^{-2}$. Extraction regions
for the shell and background are shown.
}
\end{figure}

\section{Observations and Data Reduction}

X-ray observations of Kes~75 were carried out in October 2000, for
a total of $\sim 39$~ks, and in June 2006, over four pointings with
a total exposure time of $\sim 156$~ks. Observations were made with
the Advanced CCD Imaging Spectrometer ($ACIS$) instrument on board the
$Chandra$ $X$-$ray$ $Observatory$ ($CXO$), providing a spatial resolution
of 0.5\arcsec over the range 0.5-10 keV. All data were re-processed
using the most recent calibration files, and cleaned for episodes of
high background. The total good exposure time for the five pointings was
$\sim 188$~ks. Figure 1 (bottom) shows the X-ray image displayed with
a logarithmic intensity scale, showing the both the central PWN and the
remnant shell.

Infrared observations were made on September 15, 2005, using both the
IRAC and MIPS cameras on the $SST$, and were processed and mosaiced
using standard pipelines. Images were obtained at 3.6, 4.5, 5.8, 8, and
24 $\micron$, the shorter wavelength images from the IRAC instrument
and the 24 $\micron$ image from MIPS. The pixel size is 1.2$\arcsec$
for the IRAC bands and 2.45$\arcsec$ for MIPS, roughly matched to the
(wavelength-dependent) diffraction-limited point spread function of the
telescope. For each of the IRAC bands, we obtained a 36-point dither
pattern of pointings at the source, with a 30 s integration time per
pointing, yielding a total exposure of 1080~s. For the MIPS data, we
obtained a total of 4.5 ks using a series of 30 s integrations.

The remnant is not detected in any of the IRAC bands, but the two limbs
of the shell apparent in the radio and X-ray images are clearly visible
at 24 $\micron$. Figure 1 (middle) shows the 24 $\micron$ image, overlaid
with X-ray contours to highlight the similarities in morphology. Because
Kes~75 lies in the Galactic plane, there is considerable contamination
from foreground and background emission. The central PWN is not detected
in any of the infrared bands.

To extract X-ray spectra, we used the `specextract' routine in the
CIAO software package, grouping to obtain a minimum of 25 counts/bin.
We extracted individual spectra for each observation segment using
extraction regions corresponding to the southeast (SE) and southwest
(SW) limbs and to the PWN, with a background region outside the shell
(Figure 1, bottom). We used these same regions to extract infrared
fluxes from the $SST$ images. X-ray spectral fitting (discussed in
greater detail below) was done using the XSPEC software package.

\section{Analysis}

To characterize the X-ray spectrum of the remnant shell, we fit the data
to a non-equilibrium ionization (NEI) collisional plasma plane-shock
model (an updated version of the XSPEC model `vpshock') with foreground
absorption. We fit the spectrum of each limb independently, jointly
fitting spectra from each of the observation segments. We find significant
excess at the high energy end of the spectrum, consistent with previous
fits of X-ray data for Kes~75 (H03). H03 discuss possible origins of
this hard tail, such as cosmic-ray acceleration at the shock front or
foreground dust-scattering of emission from the PWN, and add a power law
component to the model to account for this emission. We find that even
with the addition of a power law, the model fits to the spectra are poor
(with a reduced chi-squared $\chi_r^2 > 1.6$), the primary problem being
significant residuals around prominent emission lines, and in a broad
band near 1~keV.

As an alternative to the power law component, we investigated fits with a
second, higher temperature thermal component. As we describe below, this
two-component thermal model yields significantly better fits, with the
high-temperature component accommodating the hard emission. This component
requires enhanced abundances of Si, S, Ar, and Fe, suggesting a scenario
in which the low-temperature component corresponds to forward-shocked
circumstellar/interstellar material while the high-temperature component
originates from reverse-shocked ejecta.


\begin{deluxetable}{lccccc}
\tablecolumns{6}
\tabletypesize{\scriptsize}
\tablewidth{0pc}
\tablecaption{Infrared Flux Densities}
\tablehead{
\colhead{region} &
\colhead{3.6 $\micron$} &
\colhead{4.5 $\micron$} &
\colhead{5.8 $\micron$} &
\colhead{8 $\micron$} &
\colhead{24 $\micron$}
}
\startdata
Thermal shell & $<$0.04 & $<$0.05 & $<$1.92 & $<$0.72 & $18.9\pm 3.6$ \\
PWN & $<$0.068 & $<$0.061 & $<$0.044 & $<$0.062 & $<$0.66 \\
\enddata
\tablecomments{
All flux densities are extinction-corrected and quoted in Jy. The 3.6,
4.5, 5.8, and 8 $\micron$ data are taken with the IRAC instrument,
and the 24 $\micron$ data with the MIPS camera.
}
\end{deluxetable}


Modeling collisionally heated dust requires the infrared data as input,
as well as the hot electron gas density $n_e$, electron temperature
$T_e$, ion temperature $T_i$, shock timescale $\tau = n_e t_{shock}$,
and volume emission measure $EM = n_e n_H V$. We are only able to
measure directly $T_e$ from the X-ray spectrum, and we assume $T_i=T_e$.
This assumption may be incorrect, particularly given the young age of
the remnant, but we note that at temperatures of about 1 keV, electron
collisions are expected to dominate grain heating (Dwek and Arendt 1992),
and so any additional effect from a slightly higher ion temperature
would make only a marginal contribution to an increase in emission
from the dust. All these parameters, together with standard grain size
and composition distribution assumptions (Weingartner \& Draine 2001),
give a predicted infrared emission spectrum shape, the normalization of
which can be adjusted to fit the observed infrared flux. This, in turn,
gives a value of the dust-to-gas mass ratio. With this in mind, we note
that if ion temperatures are higher than electron temperatures, we will
overestimate the amount of dust present, since higher temperatures would
require less dust to produce the same emission at a given wavelength.
The reader is referred to B06 for a detailed description of the dust
modeling.

B06 also used the measured infrared flux values and the 70/24 $\micron$
flux ratio to further constrain these models and to make a prediction
about dust destruction (since a larger-than-expected 70/24 $\micron$
ratio would imply destruction of the smaller grains). However, in the
absence of data in the $70 \micron$ band, we had to rely on our $24
\micron$ detection and IRAC band upper limits for our constraints.

In order to extract infrared fluxes, we defined source regions consistent
with the emission observed in the Chandra images, excluded any obvious
point sources from the infrared data, and then calculated a total
background-subtracted flux with the ``funtools'' package in ds9, using
three different background regions for each infrared image. For the 24
$\micron$ detection, we averaged the calculated background-subtracted flux
for the three different background regions, while for the IRAC images,
we used the highest value attained from the three different backgrounds
to get conservative flux upper limits. PWN upper limits were obtained in
a similar manner. Formal errors for the infrared flux extractions were
determined from the width of histograms of pixel values in the region,
and were determined to be on the order of 10\%. The errors we quote,
however, are dominated by uncertainties in the background caused by
confusion along the line of sight, which we quantify as the largest
deviation from the mean of the three background-subtracted results.

To calculate the hot gas densities required to constrain the dust models,
we use the volume emission measure ($n_e n_H V$), taken directly from
the normalization of the XSPEC 'vpshock' model:
\begin{equation} K = 10^{-14}\frac{n_e n_H V}{4 \pi d^2} {\rm\
cm}^{-5} \end{equation} 
where $n_e$ is the electron density, $n_H$ is the gas density, $V$ is the
volume of the emitting region, and $d$ is the distance to the remnant. To
calculate the volume of the emitting regions, we take the area of the
regions indicated in Figure 1 (bottom) and assume a slab-like volume
whose depth along the line of sight is equal to the length of the long
axes of the regions. We expect this method to give an upper limit to the
volume estimate, so we introduce a filling factor $0<f<1$ to represent
the total fraction of the volume that is actually emitting; the filling
factors could be considerably smaller than 1 if there is significant
clumping in the gas. We then assume that $n_e / n_H = 1.18$, which is
true for a fully ionized gas with solar abundances, and that $d = 19$ kpc.

\section{Results}

\subsection{SNR Infrared flux}

We do not detect Kes~75 in any of the four IRAC bands, but we do find
a clear detection in the 24-$\micron$ MIPS band. This is consistent
with the B06 and W06 LMC SNR results, for which detections were only
found in the MIPS bands, and with the GLIMPSE SNR results which detected
only $\sim10$\% of known Galactic SNRs in the IRAC bands. Infrared flux
results are presented in Table 1, where we have corrected the observed
flux values using extinction corrections from Indebetouw et al. (2005).

\begin{figure}[t]
\includegraphics*[width=3.5in]{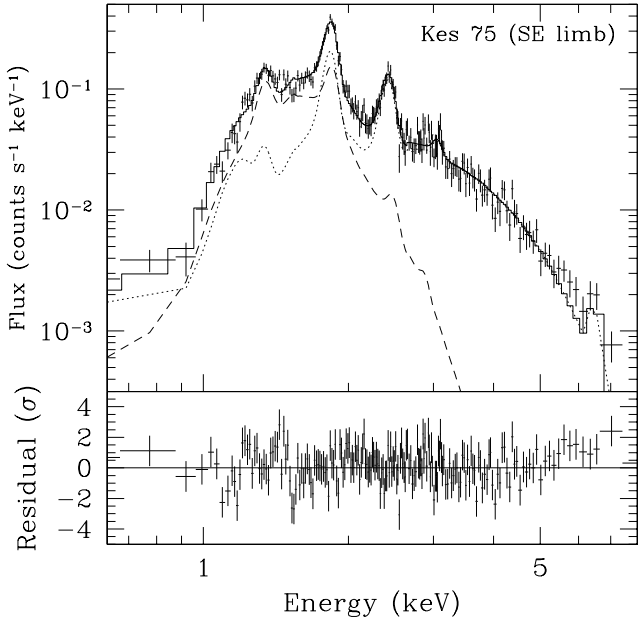}
\caption{
The spectrum of the Kes~75 thermal shell (SE limb, taken from a 54~ks
segment of {\sl Chandra} data), with a two-component thermal model. Dashed
(dotted) curves show the low (high) temperature thermal model components.
}
\end{figure}

\subsection{X-ray spectral fitting}

As a starting point to spectral fitting of the shell emission from
Kes~75, we determined the column density of the foreground gas through
fits to the PWN spectrum. We extracted spectra from the five observation
segments using a circle centered on the pulsar, with a 7~arcsec radius,
but excluding emission from the pulsar itself. An absorbed power
law provided an excellent fit to the $\sim 64\,000$-count spectrum
($\chi^2_r = 1.01$), with $\Gamma = 1.96 \pm 0.04$ and $N_H = (4.03 \pm
0.07) \times 10^{22}$ cm$^{-2}$. The $\sim 40\,000$-count spectrum from
an annulus extending from a radius of 7~arcsec to 15 arcsec, excluding a
small circle around a jet-like feature, gave best-fit values of $\Gamma
= 2.11 \pm 0.05$ and $N_H = (4.04 \pm 0.10) \times 10^{22}$ cm$^{-2}$
($\chi^2_r = 0.99$). Details of the spectrum and structure of the PWN are
the subject of a forthcoming paper. Based on the above fits, we adopt
a value of $N_H = 4 \times 10^{22}$ cm$^{-2}$ for the shell of Kes~75.
We note that this value is in excellent agreement with that derived by
H03 through fits to the PWN.

The shape of the high energy continuum (above 3~keV) for the SNR shell
is consistent with a bremsstrahlung model with $kT \sim 1.8$~keV. With
$N_H$ fixed as described above, extrapolating back to lower energies
yields a large deficit relative to the observed flux in the $1-2$~keV
band; an additional soft thermal component is required. We thus
investigated a model with two vpshock model components, one with a
temperature near that from the bremsstrahlung model, and one set lower
to accommodate the soft emission. We allowed the temperature,
ionization timescale, and normalization to vary for each of the thermal
components, and find that significant residuals remain around the lines
of Si, S, and Ar, as well as in a broad band near $\sim 1$~keV, where
Fe-L emission dominates. When the abundances of these elements are
allowed to vary (independently) in the high temperature component, we
find a much-improved fit with large enhancements above solar values. In
Figure 2 we plot the spectrum of the SE limb from one of the
observation segments. Also shown is the best-fit model, as well as each
individual emission component.  Fitting results are summarized in Table
2, where the errors on the fit parameters represent 90\% confidence
intervals.\footnote{We note that H03 find similar $\chi_r^2$ values for
an ad hoc model of bremsstrahlung emission accompanied by gaussian
lines at discrete energies, but reject this model as non-physical based
on the lack of connection between derived temperature and the centroids
and relative strengths of the lines.}


\begin{deluxetable}{lcccc}
\tablecolumns{3}
\tabletypesize{\scriptsize}
\tablewidth{0pc}
\tablecaption{Spectral Fits to the Kes~75 Thermal Shell}
\tablehead{
\colhead{Parameter} &
\colhead{SE Rim} &
\colhead{SW Rim} \\
}
\startdata
$N_H$ ($\times$10$^{22}$ cm$^{-2}$) & 4.0 (fixed) & 4.0 (fixed) \\
$kT_1$ (keV) & $0.25 \pm 0.01$ & $0.22\pm^{0.02}_{0.01}$ \\
Abundances & solar & solar \\
$\tau_1$ (s cm$^{-3}$) & $>1.5\times 10^{12}$ & $>1.5\times 10^{12}$ \\
$F_1$\tablenotemark{a} ($\times$10$^{-10}$ ergs cm$^{-2}$ s$^{-1}$) & 4.6 & 2.2 \\
\\
$kT_2$ (keV) & $1.38\pm 0.02$ & $1.50\pm^{0.98}_{0.07}$ \\
$[$Si$]$\tablenotemark{b} & $3.7\pm 0.2 $ & $3.6\pm 0.5$ \\
$[$S$]$ & $13.5 \pm 0.7$ & $10.2\pm 1.3$ \\
$[$Ar$]$ & $10.5\pm 2.3$ & $6.9\pm^{3.4}_{3.2}$ \\
$[$Fe$]$ & $14.5\pm 1.8$ & $14.2\pm^{5.2}_{4.3}$\\
$\tau_2$ (s cm$^{-3}$) & $(7.0\pm 0.3) \times 10^9$ & $(6.1\pm 0.4) \times 10^{9}$\\
$F_2$ ($\times$10$^{-10}$ ergs cm$^{-2}$ s$^{-1}$) & 2.2 & 1.0 \\
\\
$\chi^{2}_{r}$ (dof) & 1.34 (840) & 1.31 (522)\\
\enddata
\tablenotetext{a}{X-ray fluxes are unabsorbed in the 0.5 - 10 keV band.}
\tablenotetext{b}{Abundances are relative to solar values.}
\end{deluxetable}


While the X-ray spectrum from an SNR is, in reality, comprised of
emission from a distribution of temperatures, compositions, and
ionization states, we broadly associate the low-temperature component
found here with the forward-shocked CSM/ISM, and the high-temperature,
metal-enriched component with the reverse-shocked ejecta. The ionization
timescales for the CSM/ISM component are poorly constrained on the high
end, and we are only able to reliably obtain lower limits. The fits are
remarkably similar for the two limbs, particularly given the considerable
difference in X-ray/IR flux ratios observed in Figure 1. The best-fit
model for the SE limb, which has a higher X-ray flux than the SW limb,
yields a slightly higher swept-up CSM/ISM temperature and a slightly
lower ejecta temperature, but these variations are not of high statistical
significance. We do not find any evidence for spectral variations within
the limb regions, although the reduction in the number of counts in each
spectrum results in a corresponding increase in the fit uncertainties,
limiting our ability to discern small spatial variations that might be
expected from the two emission components.

\subsection{Densities and Shock Ages}

As described above, we can estimate the densities of the emission
components from the normalization provided by the X-ray model fits along
with an estimate of the emitting volume (Eq.~1). Using values from Table
2, we find filling-factor-dependent post-shock electron densities for both
the swept-up CSM/ISM component and ejecta component from each limb. The
values we get are fairly high, especially for the CSM/ISM component
(58$f_1^{-1/2}d_{19}^{-1/2}$ cm$^{-3}$ and 36$f_1^{-1/2}d_{19}^{-1/2}$
cm$^{-3}$ for the SE and SW limbs, respectively). In addition, even
with these high densities, the forward shock ages implied from the
measured ionization timescales ($>$2400 yrs) are significantly higher
than current best estimates of the pulsar age from spindown measurements
($<$900 yrs). For the reverse shock, we obtain more modest densities of
$\sim 9.6 f_2^{-1/2}$ cm$^{-3}$ ($\sim 2.2 f_2^{-1/2}$ cm$^{-3}$) for
the SE (SW) region, and an age of around 75 years, which, though young,
is not completely implausible for recently shocked ejecta. We scale
quantities with different filling factors $f_1$ and $f_2$ to indicate
the possibility that the two emission components may occupy different
fractions of the total volume (but we treat these as being the same in
the SE and SW limbs). We note that the density estimate for the ejecta
component (which comprises the bulk of the observed flux, but
a considerable minority of the unabsorbed flux) is similar to that
obtained for a single thermal component by H03. It is a factor of
$\sim 5$ higher than that obtained by Blanton \& Helfand (1996), but
this is primarily because they assumed that the emission completely
filled a shell around the PWN (since ASCA was unable to resolve the
emission and show that the thermal component was primarily from the
much smaller SE and SW limbs). 

\begin{figure}[t]
\plotone{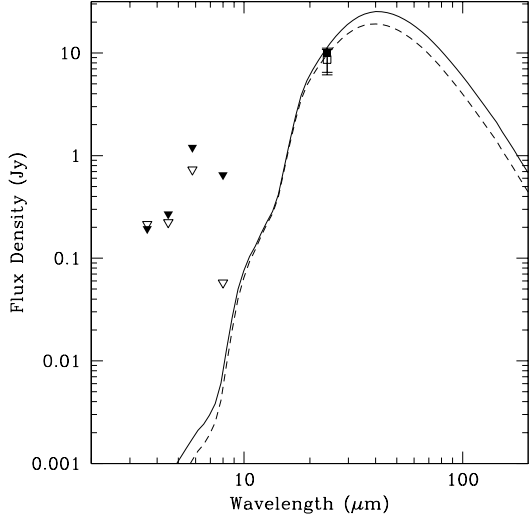}
\caption{
Dereddened IR emission from the shell of Kes 75, along with the dust-emission
model described in the text. Inverted triangles indicate upper limits.
Filled symbols and the solid curve correspond to the SW shell while
open symbols and the dashed curve correspond to the SE shell. The model
is derived from the electron temperature and density implied by the
X-ray measurements, with the $24 \mu$m measurements setting the normalization
for the hydrogen density.
}
\end{figure}

\subsection{Dust models}

The measured 24 $\mu$m fluxes of 2.4 Jy and 2.8 Jy in the SE and SW
limbs, respectively, can be used to estimate dust masses present in
the shocked ambient medium. Here we treat the IR emission as arising
exclusively from dust. This is consistent with the X-ray results that
indicate a swept-up CSM/ISM mass that is a factor of $\sim 6$ larger
than the X-ray-emitting ejecta mass.  The high ($4 \times 10^{22}$
cm$^{-2}$) column density $N_H$ toward Kes 75 implies a very high
($A_V \sim 22^{\rm m}$) optical extinction, and substantial extinction
($A_{24 \mu{\rm m}}=1.4$) even in the MIPS 24 $\mu$m band (we used
an infrared extinction curve from Chiar \& Tielens 2006 valid for the
local diffuse ISM). The dereddened 24 $\mu$m fluxes are 8.7 Jy and 10.2
Jy in the SE and SW limbs, respectively, Because the shape of the IR
spectrum is not known, we must rely on modeling in making dust mass
estimates. As described in Section 3 (and, in more detail, in B06),
the spectral shape is determined by the grain temperatures, which depend
primarily on the temperatures and densities of the X-ray emitting plasma. The
electron temperatures are 0.25 keV and 0.22 keV in the SE and SW limbs
(Table 2).  A lower limit to the electron densities, 50 cm$^{-3}$, is obtained
from limits on $\tau$ listed in Table 2 and the SNR age $t_{SNR}$ of
less than 884 yr (Livingstone et al. 2006). Electron densities derived
from emission measures depend on the unknown filling fraction $f$; by
setting $f=1$ and assuming a distance of 19~kpc, we obtain comparable
density lower limits of 58 cm$^{-3}$ and 36 cm$^{-3}$ in SE and SW
limbs, respectively. The spectral energy distribution predicted from
these inputs, and constrained by the observed $24 \mu$m flux, is plotted
in Figure 3. Triangles indicate upper limits from the IRAC data, and
squares represent the measured flux from the MIPS data.

The estimated dust masses are 0.03 M$_\odot$ and 0.05 M$_\odot$ in SE
and SW, for plane shocks with age of 884 yr, and temperatures and density
upper limits just quoted. These dust masses should be considered as upper
limits to the actual dust masses in the limbs, as less dust is required
to produce the observed fluxes at higher plasma densities (which would
be implied for filling factors smaller than the assumed value). About
40\%\ of the preshock dust mass was destroyed by sputtering in these
shock models; dust destruction rates are higher at higher densities.
In addition, IR line emission from [Fe II] or [O IV], which been observed
in N49 (Williams et al. 2006b) and has been suggested as a contributor
to the emission from 1E~0102.2-7219 (Stanimirovi\'{c} et al. 2005), may
contribute a small fraction of the flux from the ejecta component in
Kes~75. If present, this will further reduce our dust mass estimates.
We note that we have assumed a grain size distribution and composition
that is typical of the Milky Way ISM. With a detection at only one
wavelength in the IR, the data are not sufficient to probe potential
variations from these values.

\subsection{PWN spectrum}

We do not detect the central PWN in any of the $SST$ bands we observed,
but we list upper limits to the flux in Table 1, and we plot these (in
Figure 4) against current radio and X-ray data from Salter et al. (1989),
Bock and Gaensler (2005), and H03. We find that the IRAC upper limits do
not introduce stringent constraints on the spectral break frequency. While
a change in spectral index is required between the radio and X-ray bands,
with extrapolation of the radio and X-ray spectra indicating a break
somewhere near the IR band, the {\sl SST} data do not constrain where
this break occurs.

\begin{figure}[t]
\plotone{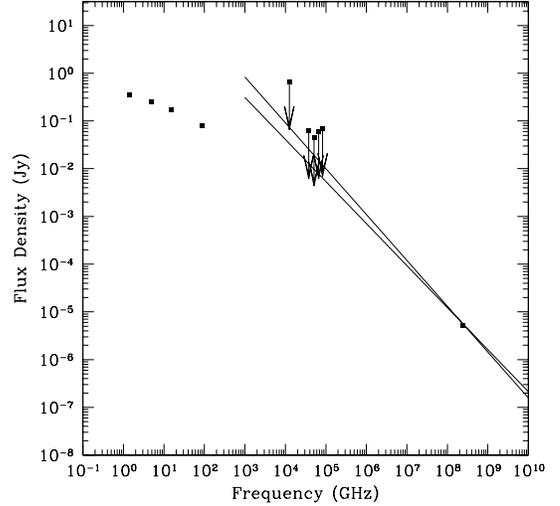}
\caption{
The PWN spectrum. Note the spectral break necessary between the radio
points (from Salter et al. 1989; Bock and Gaensler 2005) and the X-ray
regime (Helfand et al. 2003). Our infrared upper limits are
consistent with both X-ray and radio data (marginally so with X-ray at
the 8 $\micron$ point) suggesting a spectral break in or near the infrared.
}
\end{figure}

\section{Discussion}

\subsection{The remnant's shell}

Using our density estimates from both limbs to calculate the total gas
masses, we find $\sim 94f_1^{1/2}d_{19}^{5/2}$ $\msun$ for the cool
forward-shock component and $\sim 15f_2^{1/2}d_{19}^{5/2}$ $\msun$
for the ejecta component, where we have added the contributions for the
two shells. The value for the CSM/ISM component is particularly high,
especially given the large expansion velocity implied by the inferred
size and age, although clumping of the gas (i.e., a smaller filling
factor) would yield smaller values. We also note that these estimates
correspond only to the observed shell segments, which occupy roughly
20-30\% of the circumference of the remnant.

While the adopted two-component thermal model provides an adequate fit
to the X-ray spectrum of Kes~75, it is clearly only an approximation
for the forward and reverse shock emission components. We are unable
to constrain more complex distributions of the temperature, ionization
state, and abundances, particularly given the very high absorption, but
we believe that the distinct components identified in the spectral fits
provide adequate evidence to identify the presence of shocked CSM/ISM
material in the presence of enriched ejecta. We note, however, that
the high Fe abundance inferred for the ejecta component is problematic
in that it implies much higher amounts of Fe than expected for a very
massive stellar progenitor. While the uncertainties listed in Table 2
correspond to formal $\chi^2$-fitting errors, the enhanced Fe abundance
is preferred by the fits solely to reduce residuals in the broad Fe-L
region near $\sim 1$~keV, where the high absorption makes it particularly
difficult to rule out a more complex temperature distribution for the
forward shock component that could potentially reduce these residuals.

\subsection{The remnant's dust content}

The dust/gas mass ratios we find are $5 \times 10^{-4}$ and $1.4 \times
10^{-3}$ for the SE and
SW limbs, respectively.  These are considerably smaller than the value
of $\sim 0.7\%$ normally assumed for the Galaxy. The nominal Galactic
value is calculated by measuring the extinction from nearby bright stars;
using SNRs to probe dust provides a different sample of the ISM dust
content in the Galaxy, potentially affected by local grain destruction
by the SNR shock (which may represent as much as 40\% of the dust mass
in the preshock material, as described in Section 4.4). We note that
the dust mass estimates are based on the assumption that all of the IR
emission results from dust. Any contribution from line emission would
reduce the dust/gas mass ratio even further. We also note, though, that
the gas mass values are derived from the low-temperature component of
a heavily absorbed X-ray spectrum.

Our low dust/gas ratio is consistent with the results reported by B06
and W06 in the LMC, who found discrepancies of factors of $\sim 5$
between their deduced dust/gas ratios and the (lower than Galactic)
values commonly assumed for the LMC. W06 speculate that if grains are
porous, smaller dust masses could provide the observed extinctions,
while larger masses would be required to produce the observed IR emission
because of more efficient dust destruction, bringing the two types of
estimates closer together.

In order to place better constraints on the dust component, additional
infrared observations are required.  Infrared spectra would more tightly
constrain the dust models and allow us to state more confidently the
dust composition of the remnant. A spectrum would also reveal how much,
if any, of the 24 $\micron$ flux might be from emission lines (such as
[Fe II] or [O IV]) rather than emission from heated dust.

\subsection{The remnant's progenitor}

Chevalier (2005) discusses the origins and evolution of massive star
core-collapse supernovae, to which class Kes~75 belongs, and notes in
particular that Kes~75 has characteristics suggestive of an origin from
a Type Ib or Type Ic event. SN Ib/c are born from Wolf-Rayet (WR)
stars, which are characterized by high mass-loss rates, and especially
high-velocity stellar winds. These WR winds are 100-200 times faster
than winds from earlier mass-loss stages, and so they sweep up the
circumstellar medium and slow-velocity wind material, and clear out a
bubble around the star, bounded by a dense and clumpy shell extending
up to $\geq$10 pc over a WR lifetime of $\sim$$2\times 10^5$~yr. Thus,
when the star finally collapses and produces a SN Ib/c, the ejecta
travel initially through very low-density gas, and then eventually
begin to interact with this dense WR shell.

Assuming a WR picture for a progenitor for Kes~75 answers several
questions, most prominently the large size of the remnant: if ejecta
are able to expand uninhibited through a low-density medium (cleared
out by the WR wind), then they will not sweep up mass quickly, and thus
will not slow down significantly. This indication, and the young shock
age of the reverse-shocked ejecta (implying recent shocking of the WR
shell), both point to this progenitor model. This bubble model might
also help explain the high density and clumpiness that we seem to be
observing, and mixing of ejecta and swept-up material across the contact
discontinuity due to the interaction with a dense shell may explain the
lack of any obvious spatial separation of these components. Finally,
dust depletion by the WR wind could also contribute to the very low
inferred dust-to-gas ratio in Kes~75.

The X-ray luminosity for Kes~75, based on the flux values from Table~2,
is extremely high ($L_x = 4.3 \times 10^{37} d_{19}^2 {\rm\ ergs\
s}^{-1}$). Nearly 70\% of this luminosity can be attributed to the
low-temperature emission component, and this raises the concern that most
of this inferred emission is actually hidden by the high column density.
As noted above, a more complex temperature distribution for this component
could yield a significantly different unabsorbed result. However,
the luminosity for the higher temperature component is exceptionally
large as well, as is that for the pulsar and the PWN. This could
point to an error in the distance estimate of $19$~kpc. However,
a rather large error would be required. To reduce $L_{x,pwn}/\dot E$
to a value of 1\% -- similar to that for the Crab, but still a factor
of 10 larger than the (very broad) average for PWNe -- would require a
distance reduction by a factor of 4.5. This would reduce the remnant
radius and inferred free-expansion speed by the same factor, and reduce
the mass estimates by a factor of nearly 40. However, this distance
would double the already-high inferred density of the postshock gas,
and the observed column density would be anomalously high. If, instead,
the interpretation of a distant WR progenitor is correct (and we infer
this based on several lines of reasoning), then Kes~75 is the remnant
of a relatively rare event, and perhaps the fact that its properties
stand out from the rest of the population is not unexpected.

\subsection{The remnant's PWN}

Extrapolation of the radio and X-ray spectra for the PWN in Kes
75 indicate a spectral break near the IR band. The upper limits we
derive for the infrared emission are consistent with such a break,
but do not provide additional constraints because the values are
above the extrapolated break at $\nu_b \approx 5 \times 10^{14}$~Hz.
If interpreted as a synchrotron break, this would indicate a magnetic
field $B \sim 100 \mu$G given the age of the pulsar. This is somewhat
large value for a PWN, and corresponds to a magnetic energy within the
nebula that comprises nearly the entire spin-down energy deposited by
the pulsar over its estimated age. The required change in spectral index
for such a break is $\Delta \alpha \approx 0.7$, which is larger than
the change of 0.5 expected purely from synchrotron losses.  However, a
wide variety of effects can lead to spectral curvature or complex breaks,
including structure in the input particle spectrum, nonuniform magnetic
fields, and time dependence in the pulsar input power. Considerably
deeper IRAC observations could constrain the spectral behavior between
the radio and X-ray bands. Similarly, observations in the submillimeter
and TeV bands could potentially probe the mid-frequency behavior of the
spectrum as well as the high energy cut-off.

\section{Conclusions}

Using updated versions of fitting codes and new $Chandra$ observations,
we model the X-ray spectrum of Kes~75 using a two-component thermal
model in which we broadly associate a cool, solar-abundance component
with forward-shocked CSM/ISM, and a hot, metal-rich component with
reverse-shocked ejecta. The implied gas masses are somewhat high,
particularly for the cool component, suggesting that the forward shock has
encountered dense ambient material. The fits give ionization timescales
for the cool component of $\tau > 10^{12}{\rm\ s\ cm}^{-3}$, implying
an age of $> 2400$ years (older than the measured spindown age), and of
$\tau < 10^{10} {\rm\ s\ cm}^{-2}$ for the hotter, enriched component,
implying very recently shocked material ($\sim$75 years). This, combined
with the large size and high average expansion velocity for the SNR,
suggests a Wolf-Rayet progenitor that cleared out a $\sim$10 pc bubble
before exploding in a type Ib/c supernova.

We also report the detection of Kes~75 at 24 $\micron$; the morphology
of the infrared emission is spatially coincident with the observed
X-ray shell emission. We use this detection and the parameters derived
from the X-ray fits to model the dust emission of this remnant, finding
a dust/gas mass ratio of about 10$^{-3}$ or less, considerably lower than the
value of $\sim$0.7\% for the Galaxy derived from optical and ultraviolet
extinction measurements. Additional infrared observations of Kes~75
at different wavelengths would further constrain this ratio. We do not
detect the remnant at shorter wavelengths, indicating the destruction
of small grains by the shock. We suggest that the use of high-resolution
infrared imaging and X-ray spectroscopy of SNRs to constrain dust models
might provide a new sample of Galactic ISM dust measurements, distinct
from extinction measurements.

We do not detect the pulsar wind nebula in Kes~75 in any of our infrared
observations, primarily due to the high Galactic background. The upper
limits we derive are consistent with unbroken extrapolations of the both
the radio and X-ray spectra of the nebula, suggesting a spectral break
either in the infrared or between the radio and infrared.

\acknowledgments
The work presented here was supported in part by {\sl Spitzer} Grants
JPL CIT 1264892 (DJH), JPL RSA 1264893 (SPR), and JPL 1265776 (POS), as well
as Chandra Grant GO6-7053X (POS) and NASA Contract NAS8-39073 (POS).


\end{document}